\documentclass[a4paper,twocolumn,10pt]{article}
\pdfoutput=1

\usepackage[utf8]{inputenc}
\usepackage[T1]{fontenc}
\usepackage[english]{babel}
\usepackage[left=15mm, right=15mm, top=18mm, bottom=20mm, headsep=10pt]{geometry}
\usepackage{microtype}
\usepackage{fnpct}
\usepackage{titling}            
\usepackage[noblocks]{authblk}  
\usepackage{titlesec}           

\pretitle{\begin{center}\large\bf}
\posttitle{\end{center}\normalsize\vspace{-.75em}}
\preauthor{\begin{center}}
\postauthor{\end{center}}

\predate{\vspace*{-1.5em}\begin{center}\footnotesize}
\postdate{\end{center}}
\renewcommand{\thesection}{\Roman{section}}
\renewcommand{\thesubsection}{\Alph{subsection}}
\renewcommand{\thesubsubsection}{\arabic{subsubsection}}
\titleformat{\section}{\centering\bf\scshape}{\thesection.}{.75em}{}
\titleformat{\subsection}{\centering\bf}{\thesubsection.}{.75em}{}
\titleformat{\subsubsection}{\centering\it}{\thesubsubsection.}{.75em}{}
\usepackage{fancyhdr}
\usepackage[hang,flushmargin,bottom]{footmisc}
\let\oldabstract\abstract
\let\oldendabstract\endabstract
\makeatletter
\renewenvironment{abstract}
{%
               {\list{}{\addtolength{\leftmargin}{4.5em}
                        \listparindent 1.5em%
                        \itemindent    \listparindent%
                        \rightmargin   \leftmargin%
                        \parsep        \z@ \@plus\p@}%
                \item\relax}%
               {\endlist}%
\oldabstract}
{\oldendabstract}
\makeatother

\usepackage{tikz}
\usepackage{pgfplots}
\pgfplotsset{compat=newest}
\usetikzlibrary{patterns}
\usetikzlibrary{patterns.meta}
\usepgfplotslibrary{fillbetween}
\usepackage{caption}
\usepackage{subcaption}
\definecolor{red}{HTML}{DF958F}
\definecolor{green}{HTML}{A8D9A2}
\definecolor{blue}{HTML}{92BADF}
\definecolor{yellow}{HTML}{F7D67B}
\definecolor{purple}{HTML}{A894BA}

\usepackage{amsmath}
\usepackage{amssymb}
\usepackage{mathtools}
\usepackage{physics}
\usepackage{dsfont}
\usepackage{centernot}        

\let\sin\relax
\let\cos\relax
\DeclareMathOperator{\sin}{sin}
\DeclareMathOperator{\cos}{cos}

\usepackage[sort&compress,numbers]{natbib}
\usepackage[hidelinks]{hyperref}
\usepackage[capitalise]{cleveref}
\let\doi\undefined                
\usepackage{doi}                  

\begin{document}
\renewcommand{\abstractname}{\vspace{-\baselineskip}}

\title{Resource-frugal Hamiltonian eigenstate preparation\\ via repeated quantum phase estimation measurements}
\author[1]{Richard Meister}
\author[1,2]{Simon C. Benjamin}
\affil[1]{Department of Materials, University of Oxford, Oxford OX1 3PH, United Kingdom}
\affil[2]{Quantum Motion, 9 Sterling Way, London N7 9HJ, United Kingdom}
\date{\today}

\twocolumn[
\begin{@twocolumnfalse}
\maketitle
\vspace*{-2.75em}
\begin{abstract}
\noindent
The preparation of Hamiltonian eigenstates is essential for many applications in quantum computing; the efficiency with which this can be done is of key interest. A canonical approach exploits the quantum phase estimation (QPE) algorithm. We adopt ideas from variants of this method to implement a resource-frugal iterative scheme, and provide analytic bounds on the complexity (simulation time cost) for various cases of available information and tools. We propose and characterise an extension involving a modification of the target Hamiltonian to increase overall efficiency. The presented methods and bounds are then demonstrated by preparing the ground state of the Hamiltonians of LiH and H$_2$ in second quantisation; we report the performance of both ideal and noisy implementations using simulated quantum computers. Convergence is generally achieved much faster than the bounds suggest, while the qualitative features are validated.
\end{abstract}
\vspace*{2em}
\end{@twocolumnfalse}
]
\thispagestyle{empty}

\section{Introduction}

The study of complex interacting many-body systems is one area where quantum computers are anticipated to yield an advantage over classical hardware. Numerous interesting properties of such systems can be derived from a few of its eigenstates; the ground state alone is often of great significance.

To be treated on quantum hardware, the problem is usually mapped to a qubit representation. Then the key task is to prepare an eigenstate of this representation in order to examine its properties.
Typically the system of interest may have features that are already known from analytic investigation (e.g. symmetries yielding conserved quantities) or estimated either via experiments or by the use of conventional computers (e.g. approximate spectra). As discussed later, we will indeed assume that certain properties of the system may already be known through such means.

Several techniques have been proposed for the quantum computer's task of preparing an eigenstate, including ones based on eigenvalue transformations~\cite{dong2022eigenvalue}, variational approaches~\cite{peruzzo2014vqe,grimsley2019adaptive,tang2021qubit}, quantum signal processing using block-encoded Hamiltonians~\cite{lin2020nearoptimalground}, and the well-known quantum phase estimation algorithm~\cite{kitaev1995qpe}, to mention a few.

In this work, we explore methods in the spirit of quantum phase estimation and its derivatives to the effect of preparing states $\varepsilon$-close to Hamiltonian eigenstates of a system efficiently. This can be achieved in a resource-frugal fashion with a single ancilla qubit, using controlled real-time evolution, and an (ideally easy to generate) initial state that has considerable overlap with the desired target state.

The basic circuit that forms the starting point for our investigation can be seen as arising from either one of two different lines of thought. One route involves \emph{probabilistic imaginary-time evolution} (PITE)~\cite{kosugi2021pite} (see also Ref.~\cite{dong2022eigenvalue}) with its notions of superposed forward and backward time evolution. The other perspective is to view the circuit as a variant of the iterative quantum phase estimation algorithm~\cite{dobsicek2007iqpe,corcoles2021dynamic}, but with the modification that the energy of the target state has been shifted to (nearly) zero. These two lines of thought arrive at functionally identical circuits, which form the basis of our present discussion. 

The method is also very closely related to the \emph{Rodeo algorithm}~\cite{choi2021rodeo,lindgren2022rodeo}, where repeatedly time-evolving the system for random durations results in rapid convergence to the target state. For a single iteration, we utilise the same basic circuit construction as Ref~\cite{choi2021rodeo}. However, due to its stochastic nature, rigorous bounds and guarantees for the Rodeo algorithm are difficult to obtain. We circumvent this problem by proposing deterministic choices for the durations instead, which allows us to evaluate the properties of the resulting state more carefully.

For the basic method we show that it is straightforward to obtain analytic bounds for the target state fidelity after a given number of iterations, as well as for the success probability and expected cost. We also analyse the influence of algorithmic errors and gate noise on the resulting state. Moreover, we investigate a method which interpolates between a trivial Hamiltonian and the target Hamiltonian. We show that the total preparation cost can in some cases be lowered by partially preparing the ground state of an intermediate Hamiltonian before driving the evolution towards the target eigenstate. This procedure can be seen as the coarse-grained limit of a measurement-driven (i.e. Zeno effect based) adiabatic preparation (see, for example, Ref.~\cite{zhao2019measurement}).

Finally, we consider an example involving the electronic structure Hamiltonian of LiH in second quantisation, and evaluate the analytic bounds versus the results from an exactly-simulated quantum computer.

\section{Hamiltonian eigenstate preparation}

In this section, we describe the method we use to prepare Hamiltonian eigenstates in detail, provide circuit implementations, and derive bounds on the quantities of interest.

First, consider a Hamiltonian $H$ on $n$ qubits -- whose eigenstates we denote as $\ket{\varphi_j}$ with their corresponding energies being $E_j$ -- and an arbitrary initial state $\ket{\psi_0}$. The goal is to produce a sequence of states $\ket*{\psi_k}$ that converges to a desired eigenstate of the Hamiltonian $\ket{\varphi_\nu}$,
\begin{equation*}
    \lim_{k\rightarrow\infty}\ket*{\psi_k} = \ket{\varphi_\nu},
\end{equation*}
and determine an iteration index $\bar{k}$ at which it can be guaranteed that the target state infidelity falls below a given threshold
\begin{equation*}
    1 - \lvert\braket*{\varphi_\nu}{\psi_{\bar{k}}}\rvert^2 \leq \varepsilon.
\end{equation*}

\subsection{Requirements}

We assume that the following information and operators are available. First, the energy $E_\nu$ of the target state is already known to some approximation. We denote the approximate energy as $\tilde{E}$, where $E_\nu = \tilde{E} \pm \delta$, with some uncertainty $\delta \geq 0$.

Second, a lower bound $\Delta$ on the minimum difference of $E_\nu$ to any other energy $E_j$,
\begin{equation*}
    \Delta \leq \min_{j \raisebox{.5pt}{$\scriptstyle \in$} \mathcal{S}} \lvert E_\nu - E_j \rvert,
\end{equation*}
is known, where
\begin{equation*}
    \mathcal{S} \coloneqq \{j \mid j \neq \nu \text{ and } \lvert\braket{\varphi_j}{\psi_0}\rvert^2 \centernot\ll \varepsilon \}
\end{equation*}
only contains indices for which the input state $\ket{\psi_0}$ has meaningful overlap with the corresponding eigenstate.

We also assume that we have available an equivalent upper bound for the largest energy difference
\begin{equation*}
    E_\mathrm{max} \geq \max_{j \raisebox{.5pt}{$\scriptstyle \in$} \mathcal{S}} \lvert E_\nu - E_j \rvert.
\end{equation*}
For ground state preparation tasks, $\Delta$ is the spectral gap, and $E_\mathrm{max}$ is the largest occupied energy. In the worst case, $E_\mathrm{max}$ would equal twice the operator norm $\lVert H \rVert$ of the Hamiltonian.

Lastly, in addition to standard $z$-rotation and Ha\-da\-mard gates, we need to be able to apply a \emph{controlled real-time evolution} (RTE) $U(t) \coloneqq e^{-iHt}$ to the system. This obviously implies knowledge of $H$ in a form suitable for circuit-based realisation.

\subsection{Circuit implementations}

\subsubsection{Cosine propagation}

\begin{figure}[tbh]
    \centering
    \begin{subfigure}{\columnwidth}
        \centering
\providecommand{\ket}[1]{\left|#1\right\rangle}
\begin{tikzpicture}[scale=1.100000,x=1pt,y=1pt]
\filldraw[color=white] (0.000000, -7.500000) rectangle (156.000000, 22.500000);
\draw[color=black] (0.000000,15.000000) -- (147.000000,15.000000);
\draw[color=black] (147.000000,15.000000) -- (156.000000,15.000000);
\draw[color=black] (0.000000,15.000000) node[left] {$\ket*{0}$};
\draw[color=black] (0.000000,0.000000) -- (156.000000,0.000000);
\draw[color=black] (0.000000,0.000000) node[left] {$\ket*{\psi_{k-1}}$};
\draw (4.500000, -3.000000) -- (7.500000, 3.000000);
\begin{scope}
\draw[fill=white] (9.000000, 15.000000) +(-45.000000:8.485281pt and 8.485281pt) -- +(45.000000:8.485281pt and 8.485281pt) -- +(135.000000:8.485281pt and 8.485281pt) -- +(225.000000:8.485281pt and 8.485281pt) -- cycle;
\clip (9.000000, 15.000000) +(-45.000000:8.485281pt and 8.485281pt) -- +(45.000000:8.485281pt and 8.485281pt) -- +(135.000000:8.485281pt and 8.485281pt) -- +(225.000000:8.485281pt and 8.485281pt) -- cycle;
\draw (9.000000, 15.000000) node {$H$};
\end{scope}
\draw (25.000000,15.000000) -- (25.000000,0.000000);
\begin{scope}
\draw[fill=white] (25.000000, -0.000000) +(-45.000000:18.384776pt and 8.485281pt) -- +(45.000000:18.384776pt and 8.485281pt) -- +(135.000000:18.384776pt and 8.485281pt) -- +(225.000000:18.384776pt and 8.485281pt) -- cycle;
\clip (25.000000, -0.000000) +(-45.000000:18.384776pt and 8.485281pt) -- +(45.000000:18.384776pt and 8.485281pt) -- +(135.000000:18.384776pt and 8.485281pt) -- +(225.000000:18.384776pt and 8.485281pt) -- cycle;
\draw (25.000000, -0.000000) node {$U(t_k)\vphantom{U^\dagger}$};
\end{scope}
\filldraw (25.000000, 15.000000) circle(1.500000pt);
\draw (59.000000,15.000000) -- (59.000000,0.000000);
\begin{scope}
\draw[fill=white] (59.000000, -0.000000) +(-45.000000:21.213203pt and 8.485281pt) -- +(45.000000:21.213203pt and 8.485281pt) -- +(135.000000:21.213203pt and 8.485281pt) -- +(225.000000:21.213203pt and 8.485281pt) -- cycle;
\clip (59.000000, -0.000000) +(-45.000000:21.213203pt and 8.485281pt) -- +(45.000000:21.213203pt and 8.485281pt) -- +(135.000000:21.213203pt and 8.485281pt) -- +(225.000000:21.213203pt and 8.485281pt) -- cycle;
\draw (59.000000, -0.000000) node {$U(t_k)^\dagger$};
\end{scope}
\draw[fill=white] (59.000000, 15.000000) circle(1.50000pt);
\begin{scope}
\draw[fill=white] (92.000000, 15.000000) +(-45.000000:29.698485pt and 8.485281pt) -- +(45.000000:29.698485pt and 8.485281pt) -- +(135.000000:29.698485pt and 8.485281pt) -- +(225.000000:29.698485pt and 8.485281pt) -- cycle;
\clip (92.000000, 15.000000) +(-45.000000:29.698485pt and 8.485281pt) -- +(45.000000:29.698485pt and 8.485281pt) -- +(135.000000:29.698485pt and 8.485281pt) -- +(225.000000:29.698485pt and 8.485281pt) -- cycle;
\draw (92.000000, 15.000000) node {$R_z(2\tilde{E}\,t_k)$};
\end{scope}
\begin{scope}
\draw[fill=white] (125.000000, 15.000000) +(-45.000000:8.485281pt and 8.485281pt) -- +(45.000000:8.485281pt and 8.485281pt) -- +(135.000000:8.485281pt and 8.485281pt) -- +(225.000000:8.485281pt and 8.485281pt) -- cycle;
\clip (125.000000, 15.000000) +(-45.000000:8.485281pt and 8.485281pt) -- +(45.000000:8.485281pt and 8.485281pt) -- +(135.000000:8.485281pt and 8.485281pt) -- +(225.000000:8.485281pt and 8.485281pt) -- cycle;
\draw (125.000000, 15.000000) node {$H$};
\end{scope}
\draw (136.000000, -7.500000) node[text width=144pt,below,text centered] {$\ket*{\Psi_{k}}$};
\draw[dashed] (136.000000,-7.500000) -- (136.000000,22.500000);
\draw[fill=white] (141.000000, 9.000000) rectangle (153.000000, 21.000000);
\draw[very thin] (147.000000, 15.600000) arc (90:150:6.000000pt);
\draw[very thin] (147.000000, 15.600000) arc (90:30:6.000000pt);
\draw[->,>=stealth] (147.000000, 9.600000) -- +(80:10.392305pt);
\draw[color=black] (156.000000,15.000000) node[right] {$\ket*{\eta_{k}}$};
\draw[color=black] (156.000000,0.000000) node[right] {$\ket*{\psi_{k}}$};
{\pgfresetboundingbox}
{\path [use as bounding box] (-28, -20) rectangle (176,24);}
\end{tikzpicture}
        \caption{Circuit implementation $\mathcal{C}$ of the cosine-propagation using Hadamard [$H$], $z$-rotation [$R_z(\theta)$] and controlled time evolution [$U(t)$] gates.}
        \label{fig:dist_circuit}
    \end{subfigure}

    \vspace{1.7em}
    \begin{subfigure}{\columnwidth}
        \centering
\providecommand{\ket}[1]{\left|#1\right\rangle}
\begin{tikzpicture}[scale=1.100000,x=1pt,y=1pt]
\filldraw[color=white] (0.000000, -7.500000) rectangle (124.000000, 22.500000);
\draw[color=black] (0.000000,15.000000) -- (115.000000,15.000000);
\draw[color=black] (115.000000,15.000000) -- (124.000000,15.000000);
\draw[color=black] (0.000000,15.000000) node[left] {$\ket*{0}$};
\draw[color=black] (0.000000,0.000000) -- (124.000000,0.000000);
\draw[color=black] (0.000000,0.000000) node[left] {$\ket*{\psi_{k-1}^\mathrm{PE}}$};
\draw (4.500000, -3.000000) -- (7.500000, 3.000000);
\begin{scope}
\draw[fill=white] (9.000000, 15.000000) +(-45.000000:8.485281pt and 8.485281pt) -- +(45.000000:8.485281pt and 8.485281pt) -- +(135.000000:8.485281pt and 8.485281pt) -- +(225.000000:8.485281pt and 8.485281pt) -- cycle;
\clip (9.000000, 15.000000) +(-45.000000:8.485281pt and 8.485281pt) -- +(45.000000:8.485281pt and 8.485281pt) -- +(135.000000:8.485281pt and 8.485281pt) -- +(225.000000:8.485281pt and 8.485281pt) -- cycle;
\draw (9.000000, 15.000000) node {$H$};
\end{scope}
\draw (28.500000,15.000000) -- (28.500000,0.000000);
\begin{scope}
\draw[fill=white] (28.500000, -0.000000) +(-45.000000:23.334524pt and 8.485281pt) -- +(45.000000:23.334524pt and 8.485281pt) -- +(135.000000:23.334524pt and 8.485281pt) -- +(225.000000:23.334524pt and 8.485281pt) -- cycle;
\clip (28.500000, -0.000000) +(-45.000000:23.334524pt and 8.485281pt) -- +(45.000000:23.334524pt and 8.485281pt) -- +(135.000000:23.334524pt and 8.485281pt) -- +(225.000000:23.334524pt and 8.485281pt) -- cycle;
\draw (28.500000, -0.000000) node {$U(2\,t_k)$};
\end{scope}
\filldraw (28.500000, 15.000000) circle(1.500000pt);
\begin{scope}
\draw[fill=white] (61.500000, 15.000000) +(-45.000000:27.577164pt and 8.485281pt) -- +(45.000000:27.577164pt and 8.485281pt) -- +(135.000000:27.577164pt and 8.485281pt) -- +(225.000000:27.577164pt and 8.485281pt) -- cycle;
\clip (61.500000, 15.000000) +(-45.000000:27.577164pt and 8.485281pt) -- +(45.000000:27.577164pt and 8.485281pt) -- +(135.000000:27.577164pt and 8.485281pt) -- +(225.000000:27.577164pt and 8.485281pt) -- cycle;
\draw (61.500000, 15.000000) node {$P(2\tilde{E}\,t_k)$};
\end{scope}
\begin{scope}
\draw[fill=white] (93.000000, 15.000000) +(-45.000000:8.485281pt and 8.485281pt) -- +(45.000000:8.485281pt and 8.485281pt) -- +(135.000000:8.485281pt and 8.485281pt) -- +(225.000000:8.485281pt and 8.485281pt) -- cycle;
\clip (93.000000, 15.000000) +(-45.000000:8.485281pt and 8.485281pt) -- +(45.000000:8.485281pt and 8.485281pt) -- +(135.000000:8.485281pt and 8.485281pt) -- +(225.000000:8.485281pt and 8.485281pt) -- cycle;
\draw (93.000000, 15.000000) node {$H$};
\end{scope}
\draw (104.000000, -7.500000) node[text width=144pt,below,text centered] {$\ket*{\Psi_{k}^\mathrm{PE}}$};
\draw[dashed] (104.000000,-7.500000) -- (104.000000,22.500000);
\draw[fill=white] (109.000000, 9.000000) rectangle (121.000000, 21.000000);
\draw[very thin] (115.000000, 15.600000) arc (90:150:6.000000pt);
\draw[very thin] (115.000000, 15.600000) arc (90:30:6.000000pt);
\draw[->,>=stealth] (115.000000, 9.600000) -- +(80:10.392305pt);
\draw[color=black] (124.000000,15.000000) node[right] {$\ket*{\eta_{k}}$};
\draw[color=black] (124.000000,0.000000) node[right] {$\ket*{\psi_{k}^\mathrm{PE}}$};
{\pgfresetboundingbox}
{\path [use as bounding box] (-29, -22) rectangle (151,22);}
\end{tikzpicture}
        \caption{Single-qubit phase estimation circuit $\mathcal{C}^\mathrm{PE}$, using a phase shift [$P(\theta)$] gate in addition to those mentioned above.}
        \label{fig:dist_circuit_phase_est}
    \end{subfigure}
    \caption{Functionally equivalent circuits to be used in the Hamiltonian eigenstate preparation iteration.}
\end{figure}
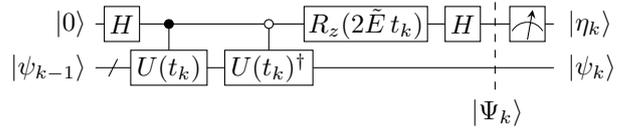
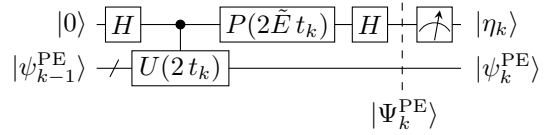

Using the established components, the preparation works as follows. We first write the initial state in terms of the Hamiltonian eigenstates
\begin{equation*}
    \ket{\psi_0} = \sum_{j=0}^{2^n-1} c_j \ket{\varphi_j},
\end{equation*}
where $c_j = \braket{\varphi_j}{\psi_0}$. After applying the circuit $\mathcal{C}$ depicted in \cref{fig:dist_circuit}, right before the measurement of the ancilla, the state in the augmented space is
\begin{multline*}
\ket{\Psi_1} = \sum_j c_j \left[\cos(\tilde{E}_j\,t_1) \ket{\varphi_j} \otimes \ket{0} \right.\\[-2.5ex]\left. {}+ i \sin(\tilde{E}_j\,t_1) \ket{\varphi_j} \otimes \ket{1}\right],
\end{multline*}
where we have introduced the notation $\tilde{E}_j \coloneqq E_j - \tilde{E}$ for the shifted energies.
The probability of measuring the ancilla qubit in state $\ket{\eta_1} = \ket{0}$, is
\begin{equation*}
    P_1 = \expval{\Pi_0}{\Psi_1} = \sum_{j} \lvert c_j \rvert^2 \cos^2(\tilde{E}_j\, t_1),
\end{equation*}
with $\Pi_0 \coloneqq \mathds{1} \otimes \ketbra{0}$ the projector onto the ancilla-zero state. After postselecting for this result, the state in the main register is
\begin{equation*}
    \ket{\psi_1} = \frac{1}{\sqrt{P_1}} \sum_j c_j \cos(\tilde{E}_j\, t_1) \ket{\varphi_j}.
\end{equation*}

After $k$ repetitions of this procedure of applying $\mathcal{C}$ and postselecting the ancilla-zero state, using different evolution times $t_\ell$, where $\ell = 1\ldots k$, we then have a total success probability of
\begin{equation} \label{eq:total_prob}
    P_k = \expval{\Pi_0}{\Psi_k} = \sum_j \lvert c_j \rvert^2 \prod_{\ell=1}^k \cos^2(\tilde{E}_j\, t_\ell)
\end{equation}
and the main register state after all operations is
\begin{equation} \label{eq:final_state}
    \ket{\psi_k} = \frac{1}{\sqrt{P_k}} \sum_j c_j \prod_{\ell=1}^k \cos(\tilde{E}_j\, t_\ell) \ket{\varphi_j}.
\end{equation}

\subsubsection{Single-bit quantum phase estimation}

An alternative -- operationally equivalent -- circuit is shown in \cref{fig:dist_circuit_phase_est}, which is essentially a single bit quantum phase estimation circuit~\cite{kitaev1995qpe} for the eigenvalue $E_\nu - \tilde{E}$. Using this circuit instead of the aforementioned cosine-propagation will yield different states
\begin{multline*}
    \ket*{\Psi_k^\mathrm{PE}} = \frac{1}{2} \left[ (\mathds{1} + e^{-2iHt_k}) \ket{\psi_{k-1}} \otimes \ket{0} \right.\\[.3ex] \left.{}+(\mathds{1} - e^{-2iHt_k}) \ket{\psi_{k-1}} \otimes \ket{1} \right],
\end{multline*}
and, starting from $\ket{\psi_0}$, after postselecting for the ancilla-zero state $k$ times, the main register contains
\begin{align}
    \ket*{\psi_k^\mathrm{PE}} &= \frac{1}{\sqrt{P_k}} \sum_j c_j \prod_{\ell=1}^k \frac{1}{2} (1 + e^{-2i\tilde{E}_j t_\ell}) \ket{\varphi_j}\nonumber\\[.75ex]
    &= \frac{1}{\sqrt{P_k}} \sum_j c_j \prod_{\ell=1}^k e^{-i\tilde{E}_j t_\ell} \cos(\tilde{E}_j t_\ell) \ket{\varphi_j}\label{eq:PE_state}
\end{align}
with the same $P_k$ as in \cref{eq:total_prob}. \Cref{eq:PE_state} differs from the states $\ket{\psi_k}$ produced by the cosine evaluation circuits only by relative phases between states $\ket{\varphi_j}$ we want to eliminate anyway -- they are therefore inconsequential to the algorithm -- and a physically irrelevant global phase of the target state $\ket{\varphi_\nu}$. The quantity of interest, $\lvert\braket{\varphi_\nu}{\psi_k}\rvert^2$, is therefore invariant to the replacement of the circuit $\mathcal{C}$ with $\mathcal{C}^\mathrm{PE}$.

For consistency and simplicity we will use $\mathcal{C}$ throughout the rest of this paper, but note that every result and proof is either directly valid or translates straightforwardly to an equivalent result for $\mathcal{C}^\mathrm{PE}$.

\subsection{Exact knowledge of \texorpdfstring{\boldmath$E_\nu$}{E\_nu}}
If the energy of the desired state is known exactly, i.e. the uncertainty $\delta = 0$, we can derive expressions for the overlap of $\ket{\psi_k}$ with the target state and the total success probability $P_k$. In this case, $\cos(\tilde{E}_\nu\, t_\ell) = \cos(0) = 1$, so \cref{eq:final_state} becomes
\begin{equation*}
    \ket{\psi_k} = \frac{1}{\sqrt{P_k}}\bigg[c_\nu \ket{\varphi_\nu} + \sum_{j\neq\nu} c_j \prod_{\ell=1}^{k}\cos(\tilde{E}_j\, t_\ell) \ket{\varphi_j}\bigg]
\end{equation*}
with the normalisation
\begin{equation} \label{eq:xi_k}
    P_k = \lvert c_\nu \rvert^2 + \underbrace{\sum_{j\neq\nu} \lvert c_j\rvert^2 \prod_{\ell=1}^q\cos^2(\tilde{E}_j\, t_\ell)}_{\eqqcolon \xi_k^2} = \lvert c_\nu\rvert^2 + \xi_k^2
\end{equation}
which is also the total success probability to measure the ancilla qubit in state $\ket{0}$ all $k$ times. This probability is bounded from below by the overlap of the initial state with the desired target
\begin{equation*}
    P_k \geq \lvert c_\nu\rvert^2 = \lvert\braket{\varphi_\nu}{\psi_0}\rvert^2,
\end{equation*}
because $\xi_k^2 \geq 0$. The overlap of the final state with the target is
\begin{equation*}
    \lvert \braket{\varphi_\nu}{\psi_k} \rvert^2 = \frac{\lvert c_\nu\rvert^2}{P_k} = \frac{\lvert c_\nu\rvert^2}{\lvert c_\nu \rvert^2 + \xi_k^2}.
\end{equation*}
If we now choose the time periods $t_\ell$ such that $\xi_k^2$ is bounded from above, we can determine how many iterations $\bar{k}$ we need to guarantee the desired target infidelity. The ideal sequence of $t_\ell$ depends on the distribution of the energies $E_j$ and the amplitudes $\lvert c_j\rvert^2$ of their states within the input state. Here we will use a generic heuristic that suppresses every energy in the interval $[\Delta, E_\mathrm{max}]$, and can thus be used even in the case of no additional information. It closely resembles a protocol the quantum phase estimation algorithm uses, albeit for slightly different reasons. However, more elaborate, tailored strategies that use additional knowledge may lead to substantially superior performance.

The longest time $t_\ell$ between measurements worth considering is $t = \pi/(2\Delta)$, as this will take the amplitude of the slowest oscillating state to exactly 0. The shortest reasonable time is $t = \pi/(2E_\mathrm{max})$, as this does the same for the fastest oscillating term. A universal heuristic is then to use the times
\begin{equation*}
    t_\ell = \frac{\pi}{2^{\bar{\ell} + 1} \Delta}
\end{equation*}
where $\bar{\ell} = (\ell - 1) \bmod N$ and
\begin{equation*}
    N = \lceil \log_2(E_\mathrm{max}/\Delta) \rceil + 1.
\end{equation*}

To quantify the convergence to the target state, we first define the maximum of the product of $N$ cosine factors as they appear in \cref{eq:xi_k} over all possible energies $\tilde{E}_j$ in the interval $\mathcal{I} \coloneqq [\Delta, E_\mathrm{max}]$ as
\begin{equation} \label{eq:gamma}
    \gamma \coloneqq \max_{\tilde{E}_j \raisebox{.5pt}{$\scriptstyle \in$} \mathcal{I}} \prod_{\ell=1}^{N} \cos^2(\tilde{E}_j\,t_\ell) \leq 1.
\end{equation}
Recall from \cref{eq:xi_k} that $\xi_k^2$ is just a weighted sum of products of the same form as in \cref{eq:gamma} with varying $\tilde{E}_j$. Thus, after each $N$ iterations, the magnitude of $\xi_k^2$ is at most a factor $\gamma$ of its previous value
\begin{equation*}
    \xi_{k+N}^2 \leq \gamma \xi_k^2,
\end{equation*}
which, when starting from $\xi_0^2$, means
\begin{equation} \label{eq:xi_bound}
    \xi_k^2 \leq \xi_0^2 \gamma^{\lfloor k/N \rfloor} = (1 - \lvert c_\nu \rvert^2) \gamma^{\lfloor k/N \rfloor},
\end{equation}
where we used that $\xi_0^2 = 1 - \lvert c_\nu \rvert^2$, because $\sum_j \lvert c_j \rvert^2 = 1$.

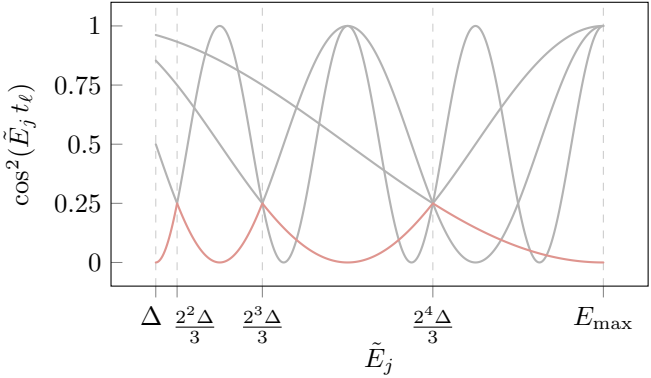
\begin{figure}[bth]
    \centering
    \begin{tikzpicture}
    \pgfplotsset{
        /pgfplots/layers/custom/.define layer set={
            axis background,pre main,axis grid,main,axis ticks,axis lines,axis tick labels,
            axis descriptions,axis foreground
        }{/pgfplots/layers/standard},
    }
    \begin{axis}[
        height=.618\columnwidth,
        width=\columnwidth,
        xlabel={$\tilde{E}_j$},
        ylabel={$\cos^2(\tilde{E}_j\,t_\ell)$},
        xlabel style={yshift=0.25cm},
        xtick={1, 4/3, 8/3, 16/3, 8},
        xticklabels={$\Delta\ $, $\ \ \ \ \frac{2^2\Delta}{3}$, $\frac{2^3\Delta}{3}$, $\frac{2^4\Delta}{3}$, $E_\mathrm{max}$},
        xtick align=outside,
        tick pos=left,
        ytick={0, .25, .5, .75, 1},
        xmajorgrids,
        major grid style={dashed, black!20},
        set layers=custom
        ]
        \addplot [red, thick, domain=1:4/3, samples=20, line cap=round] {cos(deg(pi * x / 2))^2};
        \addplot [black!30, thick, domain=4/3:8, samples=560, line cap=round] {cos(deg(pi * x / 2))^2};

        \addplot [black!30, thick, domain=1:4/3, samples=20, line cap=round] {cos(deg(pi * x / 4))^2};
        \addplot [black!30, thick, domain=8/3:8, samples=480, line cap=round] {cos(deg(pi * x / 4))^2};
        \addplot [red, thick, domain=4/3:8/3, samples=80, line cap=round] {cos(deg(pi * x / 4))^2};

        \addplot [black!30, thick, domain=1:8/3, samples=100, line cap=round] {cos(deg(pi * x / 8))^2};
        \addplot [black!30, thick, domain=16/3:8, samples=320, line cap=round] {cos(deg(pi * x / 8))^2};
        \addplot [red, thick, domain=8/3:16/3, samples=160, line cap=round] {cos(deg(pi * x / 8))^2};

        \addplot [black!30, thick, domain=1:16/3, samples=260, line cap=round] {cos(deg(pi * x / 16))^2};
        \addplot [red, thick, domain=16/3:8, samples=320, line cap=round] {cos(deg(pi * x / 16))^2};
    \end{axis}
\end{tikzpicture}
    \vspace{-1.5em}
    \caption{Illustration of the bound on $\gamma$, with the example value $E_\mathrm{max}=8\Delta$. Each oscillating line plots $\cos^2(\tilde{E}_j\,t_\ell)$ for one specific $t_\ell$ used in the algorithm. The relevant sections of each term are coloured red. The red line stays below or at $1/4$ in the entire interval, demonstrating the bound.}
    \label{fig:cosine_bound}
\end{figure}

A quite loose but intuitive upper bound on $\gamma$ can be derived as follows. Notice that when dividing $\mathcal{I}$ into sub-intervals $\mathcal{I}^{(\ell)} \coloneqq [2^{\ell}\Delta/3, 2^{\ell+1}\Delta/3]$, with $\ell = 1\ldots N$, whichever one of these intervals $\mathcal{I}^{(\ell)}$ the energy $\tilde{E}_j$ falls into, the term $\cos^2(\tilde{E}_j\,t_\ell)$ is always smaller than or equal to $1/4$,
\begin{equation*}
    \cos^2(\tilde{E}_j\, t_\ell) = \cos^2\left(\frac{\tilde{E}_j \pi}{2^\ell \Delta}\right) \leq \frac{1}{4} \hspace{.65em} \text{for }\tilde{E}_j \in \mathcal{I}^{(\ell)},
\end{equation*}
which, with a straightforward substitution of variables, is equivalent to
\begin{equation*}
    \cos^2\left(x\right) \leq \frac{1}{4} \hspace{.65em} \text{for }x \in \left[\frac{\pi}{3}, \frac{2\pi}{3}\right].
\end{equation*}
This is also illustrated in \cref{fig:cosine_bound}. We can therefore immediately see that $\gamma \leq 1/4$.

The overall convergence of the algorithm is therefore
\begin{align}
    \lvert\braket{\varphi_\nu}{\psi_k}\rvert^2 &\geq \frac{\lvert c_\nu \rvert^2}{\lvert c_\nu \rvert^2 + (1 - \lvert c_\nu \rvert^2) \gamma^{\lfloor k/N \rfloor}}\nonumber\\[.75ex]
    &\geq \frac{\lvert c_\nu \rvert^2}{\lvert c_\nu \rvert^2 + (1 - \lvert c_\nu \rvert^2) 4^{-\lfloor k/N \rfloor}}.\label{eq:bound_exact}
\end{align}
We can rearrange this expression to calculate the maximum number of required iterations $\bar{k}$ to arrive at the desired infidelity $\varepsilon \leq 1 - \lvert \braket{\varphi_\nu}{\psi_{\bar{k}}}\rvert^2$.
\begin{equation} \label{eq:kbar}
    \bar{k} = \left\lceil -\frac{N}{2} \log_2\left(\frac{\lvert c_\nu\rvert^2 \varepsilon}{(1-\varepsilon)(1-\lvert c_\nu\rvert^2)}\right) \right\rceil
\end{equation}

\subsubsection*{Implementation cost}
The cost of implementing this procedure is dominated by the real-time evolution complexity of the system. We are therefore interested in the total simulation time $\sum_\ell t_\ell$ the procedure requires. Using the sequence of $t_\ell$ described above, performing $N$ iterations necessitates time-evolving the system for
\begin{equation*}
    \sum_{\ell=1}^{N} t_\ell = \sum_{\ell=1}^{N} \frac{\pi}{2^\ell \Delta} \leq \frac{\pi}{\Delta}.
\end{equation*}

Additionally, we must account for restarts of the procedure due to a wrong measurement outcome of the ancilla qubit. Fortunately, the success probability of measuring $\ket{\eta}_k = \ket{0}$ is
\begin{equation*}
    p_k\coloneqq \frac{P_k}{P_{k-1}} = \frac{\lvert c_\nu \rvert^2 + \xi_k^2}{\lvert c_\nu \rvert^2 + \xi_{k-1}^2}
\end{equation*}
which increases quickly towards 1 -- recall that $\xi_k^2$ is exponentially decreasing -- meaning failures are most likely at the beginning of the iteration, where less cumulative simulation time has been used.

The expected total simulation time to reach and complete the $k^\text{th}$ iteration is given by sum of the simulation time to reach and pass the $k-1^\text{st}$ iteration and the cost of one additional iteration, which we must divide by $p_k$ to account for the potentially wrong measurement outcome. We can write this as a recursive function
\begin{equation} \label{eq:cost}
    \mathcal{T}(k) =
    \begin{cases}
        \frac{\mathcal{T}(k-1) + t_k}{p_k} & k > 0\\
        0 & k = 0
    \end{cases}.
\end{equation}

It is also useful to derive a bound for the maximum expected cost to reach our desired infidelity threshold of $\varepsilon$. In the most expensive case, $\xi_k^2$ only reduces after each $N$ steps, and does so by the smallest possible amount of a factor of $1/4$, as given in \cref{eq:xi_bound}. This minimal decrease of $\xi_k^2$ defers failures to later iterations in the procedure, making them more costly. This leads to the bound of
\begin{equation} \label{eq:cost_bound}
    \overline{\mathcal{T}}(k) =
    \begin{cases}
        \frac{\overline{\mathcal{T}}(k-N) + \frac{\pi}{\Delta}}{\bar{p}_k} & k > 0\\
        0 & k \leq 0
    \end{cases}
\end{equation}
with the highest possible success probability after each $N$ iterations
\begin{equation*}
    \bar{p}_k = \frac{\lvert c_\nu\rvert^2 + (1 - \lvert c_\nu\rvert^2) 4^{-\lfloor k/N \rfloor}}{\lvert c_\nu\rvert^2 + (1 - \lvert c_\nu\rvert^2) 4^{-\lfloor k/N \rfloor + 1}},
\end{equation*}
where $\overline{\mathcal{T}}(k) \geq \mathcal{T}(k)$ is the cost bound. Consequently, the total expected simulation time for preparing an eigenstate with a maximum infidelity of $\varepsilon$ from an initial state with an overlap of $\lvert c_\nu \rvert^2$ with the target is $\overline{\mathcal{T}}(\bar{k})$, with $\bar{k}$ as in \cref{eq:kbar}.

\subsection{Approximate knowledge of \texorpdfstring{\boldmath$E_\nu$}{E\_nu}}

If the energy of the desired state is not known exactly, $\delta \neq 0$, the amplitude of the target also changes over time under $\cos$-evolution, but much less so than all other states if $\delta \ll \Delta$, which we assume here. The state in the main register after $k$ iterations therefore becomes
\begin{multline*}
    \ket{\psi_k} = \frac{1}{\sqrt{P_k}}\left[c_\nu \prod_{\ell=1}^k \cos(t_\ell\,\delta) \ket{\varphi_\nu} \right.\\[-2.5ex] \left.{}+ \sum_{j\neq\nu} c_j \prod_{\ell=1}^{k}\cos(\tilde{E}_j\, t_\ell) \ket{\varphi_j}\right]
\end{multline*}
with the normalisation
\begin{equation*}
    P_k = \lvert c_\nu \rvert^2 \underbrace{\prod_{\ell=1}^k \cos^2(t_\ell\,\delta)}_{\eqqcolon\zeta_k^2} + \xi_k^2
\end{equation*}
and $\xi_k^2$ as defined above. Because we assume $t_\ell\,\delta \ll 1$, we can expand $\zeta_k^2$ into a Taylor series and truncate after the quadratic term.
\begin{multline*}
    \zeta_k^2 \approx \prod_{\ell=1}^k (1 - t_\ell^2\,\delta^2) \geq \left[\prod_{\ell=1}^N(1 - t_\ell^2\,\delta^2)\right]^{\left\lceil \frac{k}{N}\right\rceil}\\[2ex]
    \approx \Bigg[1 - \frac{\pi^2\delta^2}{\Delta^2}\underbrace{\sum_{\ell=1}^N\frac{1}{4^{\ell}}}_{\leq\frac{1}{3}}\Bigg]^{\left\lceil \frac{k}{N}\right\rceil} \geq \left[1 - \frac{\pi^2\delta^2}{3\Delta^2}\right]^{\left\lceil\frac{k}{N}\right\rceil}
\end{multline*}
Here we used the same strategy for time steps as before, and can bound the fidelity with the target state as
\begin{align} \label{eq:bound_approx}
    &\lvert\braket{\varphi_\nu}{\psi_k}\rvert^2 = \frac{\zeta_k^2\lvert c_\nu\rvert^2}{\zeta_k^2 \lvert c_\nu\rvert^2+ \xi_k^2} = \left(1 + \frac{\xi_k^2}{\zeta_k^2\lvert c_\nu\rvert^2}\right)^{-1} \\[1ex]
    &\quad= \left(1 + \frac{1 - \lvert c_\nu\rvert^2}{\lvert c_\nu\rvert^2}\ 4^{-\lfloor k/N\rfloor}\left(1-\frac{\pi^2\delta^2}{3\Delta^2}\right)^{-\lceil k/N \rceil}\right)^{-1}.\nonumber
\end{align}
Notice that the imprecise knowledge of the target energy only scales down the base of the exponential convergence by a factor, but does not limit the achievable fidelity. This reduced convergence rate translates to a slightly larger number of required iterations
\begin{equation}
    \bar{k} = \left\lceil -\frac{N\log_2\frac{\lvert c_\nu\rvert^2 \varepsilon}{(1-\varepsilon)(1-\lvert c_\nu\rvert^2)}}{2 + \log_2\left(1 - \frac{\pi^2\delta^2}{3\Delta^2}\right)} \right\rceil.
\end{equation}

For the bound of the expected cost, \cref{eq:cost_bound} remains valid, but $\bar{p}_k$ now takes the form
\begin{equation*}
    \bar{p}_k = \frac{\lvert c_\nu\rvert^2 (1 - \frac{\pi^2\,\delta^2}{3\Delta^2})^{\left\lceil k/N\right\rceil} + (1 - \lvert c_\nu\rvert^2) 4^{-\lfloor k/N \rfloor}}{\lvert c_\nu\rvert^2 (1 - \frac{\pi^2\,\delta^2}{3\Delta^2})^{\left\lceil k/N\right\rceil + 1} + (1 - \lvert c_\nu\rvert^2) 4^{-\lfloor k/N \rfloor + 1}}.
\end{equation*}

\subsection{Imperfect real-time evolution}

Except in a limited number of cases, the time evolution of a system cannot be implemented exactly, but has some algorithmic error associated with it. Without loss of generality we write the actually applied operator $U(t)$ as
\begin{equation*}
    U(t) = e^{-iHt} + \mathcal{E}(t) e^{-i\tilde{E}t}
\end{equation*}
to account for this finite accuracy.\footnote{The error operator $\mathcal{E}$ might be time-independent, but we include any possible dependence here for generality.} The resulting full space state of applying the circuit $\mathcal{C}$ is then
\begin{multline*}
    \hspace{-.8em}\ket{\Psi_1} = \sum_j c_j\left( \left[\cos([H - \tilde{E}]\,t_1) + \Re(\mathcal{E}(t_1))\right] \ket{\varphi_j} \otimes \ket{0}\right.\\[-1ex]
    \hspace{3em}\left.{}+i \left[\sin([H - \tilde{E}]\,t_1) + \Im(\mathcal{E}(t_1))\right] \ket{\varphi_j} \otimes \ket{1}\right).
\end{multline*}
As above, after $k$ iterations and post-selecting for the ancilla $\ket{\eta_k} = \ket{0}$ at every step, the main register state is
\begin{equation*}
    \ket{\psi_k} = \sum_j \frac{c_j}{\sqrt{P_k}} \prod_{\ell = 1}^k \left[\cos([H - \tilde{E}]\,t_\ell) +\Re\left(\mathcal{E}(t_\ell)\right)\right] \ket{\varphi_j}
\end{equation*}
again with the normalisation $P_k = \braket{\psi_k}$. Because the action of $\mathcal{E}$ depends on the RTE algorithm, we here establish a simple universal bound for $P_k$ and the target state fidelity $\lvert\braket{\varphi_\nu}{\psi_k}\rvert^2$. For this, we need the maximum error the RTE routine can produce\footnote{The notation $\lVert \cdot \rVert$ means the operator norm in this paper.}
\begin{equation*}
    \varepsilon_\mathrm{RTE} = \max_{t_\ell} \lVert \mathcal{E}(t_\ell) \rVert = \max_{t_\ell} \left\lVert U(t_\ell) e^{i\tilde{E}t_\ell} - e^{-i(H-\tilde{E})t_\ell} \right\rVert
\end{equation*}
and use $\Re(\mathcal{E}) = \mathcal{E}$ as the worst case situation. Assuming $\varepsilon_\mathrm{RTE} \ll 1$, we can expand $\ket{\psi_k}$ into powers of $\mathcal{E}$ and truncate after the first order. We get
\begin{equation*}
    \ket{\psi_k} = \sum_j \frac{c_j}{\sqrt{P_k}} \left[\prod_{\ell=1}^k \cos(\tilde{E}_j\, t_\ell) \ket{\varphi_j} + \sum_{\ell=1}^k \ket{\mathcal{E}_\ell}\right] + \mathcal{O}(\mathcal{E}^2)
\end{equation*}
with $\braket{\mathcal{E}_\ell} \leq \varepsilon_\mathrm{RTE}^2$. The normalisation factor can be bounded by
\begin{equation*}
    P_k = \braket{\psi_k} \leq \lvert c_\nu\rvert^2 \zeta_k^2 + \xi_k^2 + 2k\varepsilon_\mathrm{RTE} + \mathcal{O}(\varepsilon_\mathrm{RTE}^2).
\end{equation*}

Finally, the fidelity with the desired state $\ket{\varphi_\nu}$, up to order $\mathcal{O}(\varepsilon_\mathrm{RTE})$, has the bound
\begin{align}
    \lvert\braket{\varphi_\nu}{\psi_k}\rvert^2 &= \frac{1}{P_k}\left\lvert c_\nu \prod_{\ell=1}^{k}\cos(t_\ell\,\delta) + \sum_{\ell=1}^{k} \braket{\varphi_\nu}{\mathcal{E}_\ell} \right\rvert^2\nonumber\\[.5ex]
    &\geq \frac{\zeta_k^2 \lvert c_\nu \rvert^2 - 2k\varepsilon_\mathrm{RTE}}{\zeta_k^2 \lvert c_\nu\rvert^2 + \xi_k^2 + 2k\varepsilon_\mathrm{RTE}}.\label{eq:bound_RTE}
\end{align}
In contrast to the case where the target energy is not exactly known, which only scaled down the convergence rate, having an imprecise real-time evolution puts a hard ceiling on the achievable fidelity. Note, however, that this bound is extremely loose, as we assumed an adversarial error term. But, as we show in the next section, the result is qualitatively accurate.

\subsection{Gate noise}
We also consider the case of noisy quantum hardware. This may be a concern either because the algorithm is performed using physical qubits as the algorithmic qubits (as in the NISQ era), or because logical qubits are the algorithmic qubits but they are of inadequate size to guarantee a negligible total error probability (as is expected in early fault tolerant devices).

As a very simple error model, we assume depolarising noise is applied to every qubit after every gate. Depolarising noise is equivalent to a certain probability of having an unwanted and undetected Pauli operator act on the qubits. Here a \emph{gate} means a Hadamard or an exponential of a Pauli string -- sometimes called a \emph{Pauli gadget} -- of the form
\begin{equation*}
    \exp(-i\frac{\theta}{2}\bigotimes_{k=0}^{n-1} \sigma_k),
\end{equation*}
where $\sigma_k$ is a Pauli operator on qubit $k$. These Pauli gadgets occur naturally when using Trotter formulas for time evolution.

We do not derive rigorous bounds for this case, but nevertheless give a rough first-order approximation of the achievable target state fidelity with a given error rate. For this estimate, we consider the case of a large number of terms in the Hamiltonian, such that we can neglect the influence of all gates except the Pauli gadgets. As discussed above, performing $N$ iterations with the times $t_\ell$ results in a total simulation time of $\pi/\Delta$. Assuming $N_\mathrm{Trott}$ Trotter steps per unit time are required for the desired algorithmic accuracy, carrying out $N$ iterations requires
\begin{equation*}
    N_\mathrm{Pauli} = \frac{\pi L N_\mathrm{Trott}}{\Delta}
\end{equation*}
    Pauli gadgets to implement, where $L$ is the number of terms in the Hamiltonian. Because errors introduced in previous iterations are largely suppressed by the measurements in later ones, the majority of the infidelity will be caused by the last $N_\mathrm{Pauli}$ gates. If each gadget introduces an error with probability $\lambda$, the total expected fidelity of the produced density operator $\rho$ with the desired state $\ket{\varphi_\nu}$ can then be approximated by
\begin{equation} \label{eq:noisy_fidelity}
    \bra{\varphi_\nu}\rho\ket{\varphi_\nu} \approx (1 - \lambda)^{N_\mathrm{Pauli}}.
\end{equation}
Note that we do not account for the use of quantum error mitigation~\cite{cai2022quantum} which can suppress the impact of errors, typically through the use of additional repetitions, increasing the time cost.

\subsection{Morphing Hamiltonian}
Lastly, we will explore the possibility that in some cases, if the overlap of the initial state with the target is small, the total cost of the preparation can be decreased by introducing an artificial Hamiltonian
\begin{equation*}
    H_\mathrm{morph}(\alpha) = (1 - \alpha) H_\mathrm{init} + \alpha H,
\end{equation*}
with $H$ the target Hamiltonian as before, and an artificial Hamiltonian $H_\mathrm{init}$, which has the initial state $\ket{\psi_0}$ as an eigenstate. Consequently, $\tilde{E}$, $\delta$, $\Delta$, and $E_\mathrm{max}$ all become functions of $\alpha$.

This morphing Hamiltonian can then be used with a number of values $\alpha_n \in [0, 1]$, where at each value $\alpha_n$ only $N$ timesteps are performed before moving on to $\alpha_{n+1}$. At the final value of $\alpha = 1$, the full preparation is performed to the desired accuracy.

This process somewhat resembles a combination of adiabatic evolution with the quantum Zeno effect, because the state is dragged along close to the desired state by changing the Hamiltonian while simultaneously repeatedly measuring its phase change~\cite{zhao2019measurement}. We numerically investigate the coarse grained limit of this procedure with only a single intermediate value of $0 < \alpha < 1$, and demonstrate the efficacy but also limitations in \cref{sec:results}.

\section{Results} \label{sec:results}

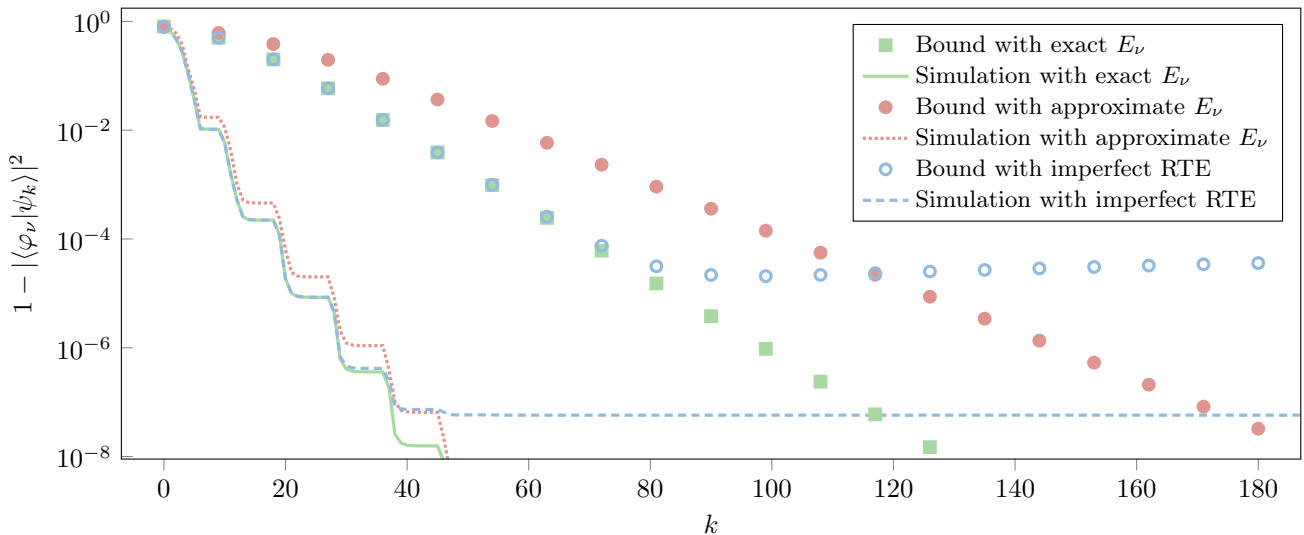
\begin{figure*}[t]
    \centering
\begin{tikzpicture}

\begin{axis}[
height=.42\textwidth,
width=.95\textwidth,
log basis y={10},
xtick align=outside,
tick pos=left,
xmin=-7, xmax=187,
ymin=9e-9, ymax=1.75,
ymode=log,
axis on top,
legend style={at={(.985,.965)},anchor=north east,font=\small},
legend cell align={left},
xlabel={$k$},
ylabel={$1 - \lvert\braket{\varphi_\nu}{\psi_k}\rvert^2$}
]
\addplot [very thick, green, mark=square*, mark size=2, only marks]
table {data/infidelity_convergence-000.dat};
\addlegendentry{Bound with exact $E_\nu$} \label{plt:bound_exact_infidelity}
\addplot [very thick, green]
table {data/infidelity_convergence-001.dat};
\addlegendentry{Simulation with exact $E_\nu$} \label{plt:exact_infidelity}
\addplot [very thick, red, mark=*, mark size=2, only marks]
table {data/infidelity_convergence-002.dat};
\addlegendentry{Bound with approximate $E_\nu$} \label{plt:bound_approx_infidelity}
\addplot [very thick, red, densely dotted]
table {data/infidelity_convergence-003.dat};
\addlegendentry{Simulation with approximate $E_\nu$} \label{plt:approx_infidelity}
\addplot [very thick, blue, mark=o, mark size=2, only marks]
table {data/infidelity_convergence-004.dat};
\addlegendentry{Bound with imperfect RTE} \label{plt:bound_rte_infidelity}
\addplot [very thick, blue, densely dashed]
table {data/infidelity_convergence-005.dat};
\addlegendentry{Simulation with imperfect RTE} \label{plt:rte_infidelity}
\end{axis}

\end{tikzpicture}
    \vspace{-.3em}
    \caption{Infidelity of the produced state $\ket{\psi_k}$ in the main register with the target state $\ket{\varphi_0}$ versus the iteration number $k$. Lines \ref{plt:exact_infidelity} \ref{plt:approx_infidelity} \ref{plt:rte_infidelity} are numerical results from preparing the ground state from the initial state in \cref{eq:init_state}, markers~\ref{plt:bound_exact_infidelity}~\ref{plt:bound_approx_infidelity}~\ref{plt:bound_rte_infidelity} are the corresponding bounds derived earlier. Different colours represent different limitations; green~\ref{plt:bound_exact_infidelity}~\ref{plt:exact_infidelity} for exact knowledge of the target state energy and perfect real-time evolution, red~\ref{plt:bound_approx_infidelity}~\ref{plt:approx_infidelity} for only approximate knowledge of the target energy but perfect RTE, and blue~\ref{plt:bound_rte_infidelity}~\ref{plt:rte_infidelity} for exact target state energy knowledge but imperfect RTE. Note that the bound and simulation of the imperfect RTE data use different errors $\varepsilon_\mathrm{RTE}$.}
    \label{fig:infidelity_convergence}
\end{figure*}

To demonstrate our algorithm numerically, we considered the Hamiltonian of LiH in second quantisation, generated using \texttt{openfermion}~\cite{mcclean2020openfermion}, and simulated its dynamics using exact quantum emulation software. The system consists of 12 qubits, has a spectral gap\footnote{We use Hartree atomic units throughout.} of $\Delta\approx 0.075$ and a maximum energy of $E_\mathrm{max} \approx 9.753$, resulting in $N = 9$ different times $t_\ell$. Starting from an initial state
\begin{equation} \label{eq:init_state}
    \ket{\psi_0} = \frac{1}{\sqrt{5}} \ket{\varphi_0} + \frac{1}{N_\mathrm{s}} \sum_{j=1}^{2^n-1} e^{-(E_j - E_0)} \ket{\varphi_j},
\end{equation}
with an appropriate normalisation factor $N_\mathrm{s}$, we executed our algorithm to amplify the ground state $\ket{\varphi_0}$ under each discussed limitation. The results of the numerical simulations together with the established bounds are shown in \cref{fig:infidelity_convergence}.

\paragraph{Exact \texorpdfstring{\boldmath$E_\nu$}{E\_nu}} The calculations using perfect knowledge of the target state energy show the expected behaviour of overall exponential convergence after some initial iterations. The step-like structure of the numerical simulation is caused by the periodic choice of simulation times. Each of those steps corresponds to a full sequence of $N$ different times $t_\ell$. Due to the rounding in the expressions for the bounds in \cref{eq:bound_exact,eq:bound_approx,eq:bound_RTE}, these also show such step-like behaviour. However, for readability we only plot every $N^\mathrm{th}$ data point for them, i.e. the bottom left corner of each step.

\paragraph{Approximate \texorpdfstring{\boldmath$E_\nu$}{E\_nu}} To show the effect of only knowing the energy of the target state approximately, we used a relatively large offset $\delta = \Delta/3$. The graph of the bound nicely illustrates the scaling down of the basis of the exponential convergence, i.e. a shallowing of the slope. The numerical simulation also shows a slightly reduced rate of convergence compared to the case of exactly known energy. We note that this is not always necessarily the case. Depending on the exact distribution of energy levels and their occupation in the initial state, either one may converge faster than the other. However, the \emph{guaranteed} convergence is always quicker the more precisely the energy of the target state is known.

\paragraph{Imperfect RTE} We also performed a calculation of the same system with precisely known target state energy ($\delta = 0$), but using a first-order product formula~\cite{trotter1959product,suzuki1976generalized} as the RTE routine, dividing the shortest time interval into 128 slices. Importantly, in the plot the bound \ref{plt:bound_rte_infidelity} and the numerical simulation \ref{plt:rte_infidelity} do not use the same $\varepsilon_\mathrm{RTE}$, because the bound is very loose. The simulation has the numerically obtained value of $\varepsilon_\mathrm{RTE} \approx 5.6 \cdot 10^{-4}$, while the bound uses the much smaller $\varepsilon_\mathrm{RTE} = 10^{-8}$. Therefore, the simulation and the bound are only qualitatively related. We still see the same pattern emerge in both cases. There is a close match between the exact simulation and the imperfect RTE solution, until some lower threshold of the infidelity is reached, after which the approximate version becomes roughly constant,\footnote{The bound even slightly increases due to some of the approximations made in its derivation.} and no further progress can be made. It is therefore evident that while imprecise energy knowledge only slows down the convergence of the algorithm, the presence of simulation errors puts a hard lower limit on the attainable infidelity.

\begin{figure}[b]
    \centering
\begin{tikzpicture}

\pgfplotsset{
    /pgfplots/layers/custom/.define layer set={
        axis background,axis grid,main,axis ticks,axis lines,axis tick labels,
        axis descriptions,axis foreground
    }{/pgfplots/layers/standard},
}

\begin{axis}[
height=.618\columnwidth,
width=\columnwidth,
log basis y={10},
xtick align=outside,
tick pos=left,
xmin=-2, xmax=32,
ymin=1.4e-9, ymax=3,
ymode=log,
legend style={at={(.985,.5)},anchor=east,font=\small},
legend cell align={left},
xlabel={$k$},
ylabel={$1 - \lvert\braket{\varphi_\nu}{\psi_k}\rvert^2$},
ytick={1e-1, 1e-3, 1e-5, 1e-7},
extra y ticks={0.0133678526414308, 0.00053763542764218, 2.15055705323763e-05, 8.60158601456718e-07, 3.43790790280352e-08},
extra y tick labels={},
extra y tick style={
    tick style={draw=none},
    grid=major,
    major grid style={gray!85, thin, densely dashed}},
set layers=custom
]
\addlegendimage{empty legend}
\addlegendentry{$\lambda$}
\addplot [thick, blue]
table {data/noisy-001.dat};
\addlegendentry{$10^{-6}$}
\addplot [thick, green]
table {data/noisy-002.dat};
\addlegendentry{$10^{-8}$}
\addplot [thick, red]
table {data/noisy-003.dat};
\addlegendentry{$10^{-10}$}
\addplot [thick, yellow]
table {data/noisy-004.dat};
\addlegendentry{$10^{-12}$}
\addplot [thick, purple]
table {data/noisy-005.dat};
\addlegendentry{$10^{-14}$}
\addplot [thick, gray]
table {data/noisy-006.dat};
\addlegendentry{$\hphantom{1}0\vphantom{0^1}$}
\end{axis}

\end{tikzpicture}
    \vspace{-1.7em}
    \caption{Infidelity of the prepared density operator $\rho_k$ with the desired target state $\ket{\varphi_\nu}$ versus the iteration number $k$ for different error rates $\lambda$. Coloured lines are numerical results, gray dashed lines show the expected approximate limit according to \cref{eq:noisy_fidelity}.}
    \label{fig:noisy_convergence}
\end{figure}
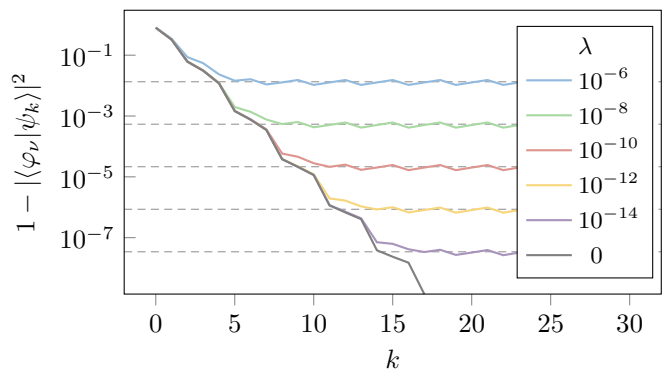

\paragraph{Gate noise} In order to demonstrate the noise resilience of the discussed method, we performed the state preparation using first order Trotterisation and different noise strengths $\lambda$. Due to the increased computational demand of using the density matrix formalism to include noisy channels, \cref{fig:noisy_convergence} shows the results for the second quantised Hamiltonian of H$_2$, a much smaller system than LiH. We see the same limiting pattern as for the case of algorithmic errors, where the state quickly converges to the desired target, but then encounters a ceiling in the fidelity caused by the errors. Our numerical results show good agreement with the approximation derived earlier in \cref{eq:noisy_fidelity}.

Notice that while a rather small error rate $\lambda$ is needed in order to obtain good fidelity, suitable quantum error mitigation (QEM) techniques can boost performance in return for additional simulation cost. An example is symmetry verification. We do not explore this further, since it is a broad topic. Moreover, the appropriate QEM will depend on multiple aspects of the task and the hardware imperfections; we refer the reader to a recent review~\cite{cai2022quantum}.

\paragraph{Morphing Hamiltonian}
To demonstrate how morphing the Hamiltonian can sometimes decrease the total cost, we again considered the LiH Hamiltonian and chose the computational basis state\footnote{Contrary to our earlier description, this is not the computational basis state with the highest possible overlap with $\ket{\varphi_0}$, but we use it as a more instructive example.} $\ket{\psi_0} = \ket{10000100011}$ as the initial state, which has an overlap with the ground state of $\lvert\braket{\psi_0}{\varphi_0}\rvert^2\approx 4.6 \cdot 10^{-3}$. The corresponding Hamiltonian we used was
\begin{equation*}
    H_\mathrm{init} = \Delta \sum_{j=0}^{n-1} (2 \psi_0^{(j)} - 1) \sigma^{z}_j
\end{equation*}
where $\psi_0^{(j)} \in \{0, 1\}$ refers to the value of qubit $j$ in the computational basis state $\ket{\psi_0}$. This choice guarantees that $\ket{\psi_0}$ is a gapped ground state of $H_\mathrm{init}$, with a gap matching that of $H$.

We note that the cost to implement the time evolution of $H(\alpha)$ will most likely be a function of $\alpha$. The exact form of this dependence will vary with the simulation technique, though the complexity of simulating $H + H_\mathrm{init}$ may be used as a cost bound for most methods. We do not explicitly address such a dependence and only report the total required simulation time in the system, regardless of the value of $\alpha$.

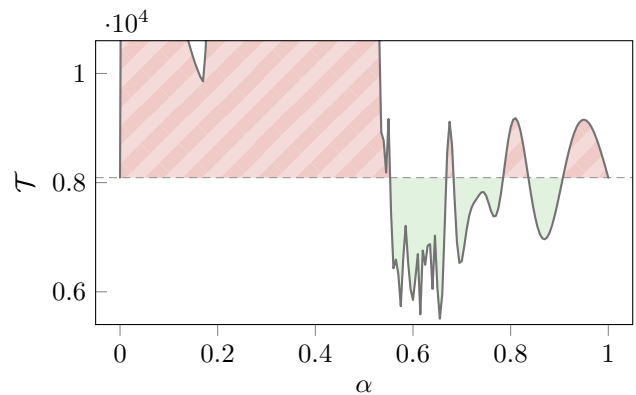
\begin{figure}[thbp]
    \centering
\begin{tikzpicture}
\usepgfplotslibrary{fillbetween}
\usetikzlibrary{patterns}
\usetikzlibrary{patterns.meta}

\pgfplotsset{
    /pgfplots/layers/custom/.define layer set={
        axis background,pre main,axis grid,main,axis ticks,axis lines,axis tick labels,
        axis descriptions,axis foreground
    }{/pgfplots/layers/standard},
}

\begin{axis}[
height=.618\columnwidth,
width=\columnwidth,
xtick align=outside,
tick pos=left,
xmin=-0.05, xmax=1.05,
ymin=5400, ymax=10600,
xlabel={$\alpha$},
ylabel={$\mathcal{T}$},
extra y ticks={8091.33282048781},
extra y tick labels={},
extra y tick style={
    tick style={draw=none},
    grid=major,
    major grid style={gray!85, thin, densely dashed}},
set layers=custom
]
\addplot [thick, black!55, name path=data, line join=round, line cap=round]
table {data/zeno-000.dat}; \label{plt:zeno_dat}
\path [name path=canonical] (\pgfkeysvalueof{/pgfplots/xmin},8091.33282048781) -- (\pgfkeysvalueof{/pgfplots/xmax},8091.33282048781);

\addplot fill between [of=data and canonical, soft clip={domain=0:1}, split,
                       every odd segment/.style={green!35},
                       every even segment/.style={preaction={fill, red!35},pattern={Lines[angle=45,distance={10pt},line width={5pt},xshift=1pt]}, pattern color=red!50}];

\addlegendimage{area legend, preaction={fill, red!35},pattern={Lines[angle=45,distance={10pt},line width={5pt},xshift=1pt]}, pattern color=red!50, draw=black!55, thick} \label{plt:zeno_bad}
\addlegendimage{area legend, fill=green!35, draw=black!55, thick} \label{plt:zeno_good}
\addlegendimage{gray!85, thin, densely dashed} \label{plt:zeno_canonical}

\end{axis}

\end{tikzpicture}
    \caption{Total simulation time cost $\mathcal{T}$ when using a single intermediate value of $\alpha$ between $H_\mathrm{init}$ and $H$, depending on the choice of $\alpha$. The cost is the mean total required evolution time and includes restarts after failed ancilla measurements. The solid line~\ref{plt:zeno_dat} represents the morphing Hamiltonian; for reference, the gray dashed line~\ref{plt:zeno_canonical} is the cost of directly preparing the ground state of $H$ without an $H_\mathrm{morph}$. In the red striped region~\ref{plt:zeno_bad}, a morphing Hamiltonian as described above increases the cost, while the in the green region~\ref{plt:zeno_good} the morphing method is cheaper.}
    \label{fig:zeno}
\end{figure}

For the purposes of a first exploration, we only considered one intermediate step between $\alpha = 0$ and $\alpha = 1$. The question of what this intermediate value should ideally be, turns out to be quite complex. \Cref{fig:zeno} shows the total required simulation time to reach an infidelity of $10^{-8}$, depending on where the intermediate $\alpha$ is placed. Green shaded regions where the graph is below the dashed line indicate values where the morphing approach is advantageous. The potentially complex behaviour of the preparation cost is exemplified in the region around $\alpha \approx 0.6$. We have identified that the rapidly oscillating character is related to the low-lying energy spectrum of $H(\alpha)$ in that area, which consists of a gap with multiple closely spaced excited states right above it.

When considering the simple case of a single $\alpha$ and a cheaply simulated Hamiltonian, the task of finding a near-optimal value is easily solved numerically. However, the general case of multiple intermediate values $\alpha_k$ remains difficult due to the rapidly increasing size of the configuration space and the non-convexity of the cost. We leave this question open for further research in the future.

\section{Discussion and Outlook}
In this paper we investigated the repeated use of a circuit closely resembling that of iterative phase estimation~\cite{dobsicek2007iqpe,corcoles2021dynamic} and the Rodeo algorithm~\cite{choi2021rodeo,lindgren2022rodeo} in order to prepare eigenstates of a Hamiltonian system from arbitrary initial states. The only required knowledge is the (approximate) energy of the target state, a lower bound of the energy gap of the target state to the closest lying occupied state, and an upper bound of the largest energy gap from the target to any other occupied state. The necessary tools to implement the presented algorithm are single-qubit gates on one ancilla qubit, as well as controlled real-time evolution (RTE) of the system.

We derived analytic bounds for the fidelity of the produced state with the target and the expected total required RTE duration for different cases. Imprecise knowledge of the target state energy results in a slower convergence rate, but does not limit the achievable fidelity. Algorithmic and gate noise, on the other hand, hardly influence the rate of convergence, but put a ceiling on how precisely the target state can be prepared.

In all cases we found asymptotically exponential convergence of the fidelity with the real-time evolution time. We also gave explicit expressions for strict bounds which are useful in practical applications, because a certain fidelity can be guaranteed after a number of iterations without the need for expensive verification.

We anticipated that our bounds would prove loose versus a specific implementation, and this indeed proved to the the case; nevertheless analytic expressions are valuable when one wishes to use a method such as this as a component of a larger algorithm, and it is desirable to bound the costs of that procedure.

The actual gate- and/or query complexity is determined by the chosen method for the controlled time evolution. For example, Hamiltonian simulation by quantum signal processing~\cite{low2017qsp} only requires $\mathcal{O}(t - \log{\varepsilon_\mathrm{RTE}})$ gates to implement the required real-time evolution, making the actual cost of the state preparation logarithmic in the desired infidelity.

Finally, we also considered a variation of the preparation process where the Hamiltonian is morphed from a trivial one to the Hamiltonian of the system of interest. We found numerically -- using the LiH system again -- that for some choices of morphing schedule, this process can decrease the cost of preparation. However, care must be taken, as unfavourable choices may easily increase the total cost. Finding a generic method to generate an efficient schedule might be an interesting topic for future research.

\section*{Acknowledgements}
The authors thank Hans Chan and Bálint Koczor for helpful discussions.
The authors would like to acknowledge the use of the University of Oxford Advanced Research Computing (ARC) facility~\cite{arc} in carrying out this work.
SCB acknowledges support from the EPSRC QCS Hub grant under agreement No. EP/T001062/1, and from the IARPA funded LogiQ project.

\bibliography{references}

\appendix
\renewcommand{\thesection}{\Alph{section}}
\renewcommand{\thesubsection}{\arabic{subsection}}
\renewcommand{\thesubsubsection}{\arabic{subsubsection}}
\titleformat{\section}[block]{\centering\bf}{Appendix \thesection:}{.75em}{}

\end{document}